\documentclass [aps,pra,twocolumn,superscriptaddress]{revtex4}
\usepackage{amsmath}
\usepackage{amssymb}
\usepackage{graphicx}
\usepackage[T1]{fontenc}
\usepackage{color}

\usepackage{soul}

\begin{document}

\title{Role of diffusive surface scattering in nonlocal plasmonics}

\author{M. K. Svendsen}
\affiliation{Department of Physics, Technical University of Denmark, DK-2800~Kongens~Lyngby, Denmark}

\author{C. Wolff}
\affiliation{Center for Nano Optics, University of Southern Denmark, Campusvej 55, DK-5230 Odense M, Denmark}

\author{A.-P. Jauho}
\affiliation{Department of Physics, Technical University of Denmark, DK-2800~Kongens~Lyngby, Denmark}
\affiliation{Center for Nanostructured Graphene, Technical University of Denmark, DK-2800 Kongens Lyngby, Denmark}

\author{N. A. Mortensen}
\affiliation{Center for Nano Optics, University of Southern Denmark, Campusvej 55, DK-5230 Odense M, Denmark}
\affiliation{Center for Nanostructured Graphene, Technical University of Denmark, DK-2800 Kongens Lyngby, Denmark}
\affiliation{Danish Institute for Advanced Study, University of Southern Denmark, Campusvej 55, DK-5230 Odense M, Denmark}

\author{C. Tserkezis}
\email{ct@mci.sdu.dk}
\affiliation{Center for Nano Optics, University of Southern Denmark, Campusvej 55, DK-5230 Odense M, Denmark}

\begin{abstract}
The recent generalised nonlocal optical response (GNOR)
theory for plasmonics is analysed, and its main input
parameter, namely the complex hydrodynamic convection-diffusion
constant, is quantified in terms of enhanced Landau damping
due to diffusive surface scattering of electrons at the
surface of the metal. GNOR has been successful in describing
plasmon damping effects, in addition to the frequency shifts
originating from induced-charge screening, through a
phenomenological electron diffusion term implemented into
the traditional hydrodynamic Drude model of nonlocal plasmonics.
Nevertheless, its microscopic derivation and justification is
still missing. Here we discuss how the inclusion of a diffusion-like
term in standard hydrodynamics can serve as an efficient vehicle
to describe Landau damping without resorting to computationally
demanding quantum-mechanical calculations, and establish a direct
link between this term and the Feibelman $d$ parameter for the
centroid of charge. Our approach provides a recipe to connect
the phenomenological fundamental GNOR parameter to a
frequency-dependent microscopic surface-response function.
We therefore tackle one of the principal limitations of the
model, and further elucidate its range of validity and
limitations, thus facilitating its proper application in
the framework of nonclassical plasmonics.
\end{abstract}

\maketitle

\section{Introduction}

Increasing interest in spatial dispersion and the nonlocal response of
plasmonic nanostructures is being observed in recent years, mainly due
to its relevance for quantum plasmonics~\cite{Tame:2013,Bozhevolnyi:2016a,
Bozhevolnyi:2016b,Zhu:2016,Fitzgerald:2016,Yannopapas:2016,Amendola:2017,
Fernandez-Dominguez:2018,Ciraci:2019} in a large number of
experimentally available plasmonic architectures with ultrafine geometrical
details~\cite{Savage:2012,Duan:2012,Scholl:2012,Scholl:2013,Wiener:2013,
Jung:2015,Raza:2015c,Krasavin:2016,Shen:2017}. Hydrodynamic descriptions
of the induced charges have been particularly emphasised~\cite{Abajo:2008,
McMahon:2010b,Raza:2011,Ciraci:2012,David:2014,Christensen:2014,Raza:2015a,
Eremin:2018,Kupresak:2020}, and widely applied in situations where small
particles or few-nm particle separations are involved. Relevant fully-quantum
mechanical studies have shown that such models can in some occasions lead to
qualitatively wrong conclusions, as, in their traditional implementation,
they neglect electron spill-out~\cite{Esteban:2012,Stella:2013,Teperik:2013a,
Yan:2015a,Hohenester:2016}, an issue which is resolved once the hydrodynamic
description is itself self-consistent~\cite{Toscano:2015,Yan:2015b,Ciraci:2016}.
Nevertheless, despite these limitations, the hydrodynamic Drude model (HDM)
still captures both qualitatively and to a large extent also quantitatively
\cite{Raza:2013a} the effect of nonlocal screening (due to spill-in promoted
by a large work function) in noble metal nanostructures, and thus provides
important insight into a wide range of experiments where spill-out and tunnelling
can be safely disregarded, within a relatively simple and computationally efficient
description. What it cannot capture, however, is the experimentally
observed~\cite{Kreibig:1970}, but theoretically challenging to quantify~\cite{Kraus:1983},
broadening of the plasmonic modes with decreasing nanoparticle (NP) size. It is
therefore useful to explore generalisations of HDM that enrich the physical description
it provides.

One of the most successful extensions to the hydrodynamic description
was achieved by the generalised nonlocal optical response (GNOR)
theory~\cite{Mortensen:2014}, which is based on a phenomenological
inclusion of diffusion of free electrons in the bulk of the metal, to
account for a series of experimentally observed plasmon damping mechanisms,
including reduction of the electron mean-free path, quantum confinement
and particularly Landau damping~\cite{Apell:1984,Ouyang:1992,Baida:2009,
Lerme:2010,Monreal:2013,Li:2013,Khurgin:2015a}. These effects are reproduced
within the GNOR theory by adding a classical diffusion term, introduced
through a drift-diffusion equation, in the hydrodynamic description of
free electrons, thus relaxing the necessity to resort to more computationally
demanding quantum-mechanical descriptions~\cite{Uskov:2014,Kirakosyan:2016,
Shahbazyan:2016,Lerme:2017}. This formalism relies on a transport
equation, established long ago for classical systems with stochastic
scattering~\cite{Chandrasekhar:1943}, i.e., convection-diffusion
theory. The importance of both convection and diffusion for nonlocal
electrodynamics was in fact predicted many years ago (see discussion
in~\cite{Landau-Lifshitz-Pitaevskii}). Diffusion has also been used
as an abstract mathematical model to explain spatial dispersion~\cite{Hanson:2010},
without however providing the physical link between diffusion dynamics and
nonlocal plasmonic broadening. An early hint of how they can be connected
through the fluctuation-dissipation theorem was given in \cite{Iwamoto:1984}.
Despite these early estimations of the importance of diffusion, however, a
more microscopic justification for its inclusion in nonlocal plasmonics in
general, and in the GNOR theory in particular, is still missing, and
the ambiguity in the choice of an appropriate hydrodynamic parameter
restrains further extension of the applicability of such models. 

Here we provide a microscopic foundation for the GNOR theory, following a
procedure of gradually increasing complexity and getting deeper into the fine
details of electron motion in a confined volume. Using the Boltzmann
equation as our starting point, we first derive a diffusion correction to
the continuity equation by assuming a weakly inhomogeneous initial electron
density. We show that convection and diffusion are both manifestations of
the same physical behaviour in the bulk, becoming important at different
frequency limits, and diffusion is practically negligible at optical
frequencies. To then justify the appearance of a diffusive term in the
GNOR model, we turn to surface-scattering and Landau damping, and show
how the GNOR diffusion constant relates to these effects. We derive a
connection between the complex convection-diffusion hydrodynamic
parameter and the Feibelman \textit{d} parameters for the centroid of
induced charge~\cite{Feibelman:1982}, which quantifies the former through
microscopic arguments. This approach is shown to provide a good agreement
with the phenomenological size-dependent broadening (SDB) correction
developed by Kreibig~\emph{et al.}~\cite{Kreibig:1970,Kreibig:1985} in
the case of a small metallic nanosphere, and it leads to reasonable
broadening for other particle shapes, for which a simple SDB
recipe is not available. Our procedure, which directly implements
information retrieved from quantum-mechanical calculations, justifies and
quantifies the phenomenological GNOR model, thus opening the pathway to 
its more widespread implementation in nonlocal/nonclassical plasmonics.

\section{Theoretical background}

Throughout the paper we focus on the intraband response of free
carriers, disregarding interband transitions. To begin with,
Maxwell's equations can be combined to obtain the wave equation
\begin{equation}\label{Eq:Maxwell}
{\boldsymbol \nabla} \times {\boldsymbol \nabla} \times \mathbf{E} =
\left( \frac{\omega}{c} \right)^{2} \mathbf{E} +
\mathrm{i} \omega \mu_{0} \mathbf{J} (\mathbf{E})~,
\end{equation}
where $\omega$ is the angular frequency of light, and $c = 1/
\sqrt{\varepsilon_{0} \mu_{0}}$ is the speed of light in vacuum,
with $\varepsilon_{0}$ and $\mu_{0}$ representing the vacuum
permittivity and permeability, respectively. The electrodynamic
response of matter is contained in the constitutive relation
between the current density $\mathbf{J}$ and the applied electric
field $\mathbf{E}$, $\mathbf{J}(\mathbf{E})$. Within linear-response
theory, the common local-response approximation (LRA) gives simply
$\mathbf{J} (\mathbf{r}) \simeq \sigma (\mathbf{r}) \mathbf{E}
(\mathbf{r})$ (Ohm's law), where $\sigma (\mathbf{r})$ is the
material conductivity, while in a nonlocal description we have
\cite{Raza:2015a} 
\begin{equation}\label{Eq:Ohm-nonlocal}
\mathbf{J} (\mathbf{r}) =
\int \mathrm{d} \mathbf{r}' \sigma(\mathbf{r}, \mathbf{r}')
\mathbf{E} (\mathbf{r}')~.
\end{equation}
This expression states the fact that the induced current density
at a point $\mathbf{r}$ inside the material should depend on the
applied electric field at all neighbouring points $\mathbf{r}'$. To
proceed we need a theoretical model for $\mathbf{J}(\mathbf{E})$, and
we will consider a hierarchy of increasing complexity, starting with
the hydrodynamic drift-diffusion theory. The hydrodynamic approach is
then justified through the Boltzmann transport equation, while
additional insight from a microscopic account of surface scattering
is provided for the diffusion term.

The hydrodynamic description relies on a classical equation of motion
for an electron in an electromagnetic field, while accounting for
quantum-pressure effects and multiple electron scattering in a
semi-classical way. In the Boltzmann approach, quantum effects of the
electron gas enter through scattering-matrix elements, where the
non-convective, random velocity components are linked to classical
diffusion. Finally, in the surface-scattering microscopic approach,
first-principles surface-response functions, namely the Feibelman
\textit{d} parameters, are introduced to obtain a connection between
the semi-classical convection and diffusion constants and the centroid
of the induced charge.

\section{Hydrodynamic approach}

Let us first briefly describe the procedure followed in the original
introduction of GNOR theory~\cite{Mortensen:2014}. We start with the
linearised hydrodynamic equation of motion for an electron in an
electric field~\cite{Raza:2011,Boardman:1982a,Tokatly:1999}
\begin{equation}\label{Eq:eom1}
\frac{\partial}{\partial t} \boldsymbol{\nu} =
-\gamma \boldsymbol{\nu} +
\frac{(-e)}{m} \mathbf{E} -
\frac{\beta^{2}}{n_{0}} \boldsymbol{\nabla} n_{1} ~,
\end{equation}
where $e$ is the elementary charge, $m$ is the electron mass, $\boldsymbol{\nu}$
is the non-equilibrium velocity correction to the static sea of electrons,
$\gamma$ is the Drude damping rate also appearing in LRA (equal to the
inverse of the electron relaxation time), while $n_{0}$ is the
equilibrium electron density (which we assume to be constant) and
$n_{1}$ denotes a small deviation from equilibrium, so that
$n = n_{0} + n_{1}$ is the total electron density. The right-hand
side contains a semi-classical correction where the pressure term
$\boldsymbol{\nabla} n_{1}$ is classical in spirit, while its strength
originates from a quantum description of pressure effects associated
with the compressible electron gas. In other words, $\beta$ is a
characteristic velocity for pressure waves associated with the finite
compressibility of the electron gas. In the high-frequency limit,
$\omega \gg \gamma$, Thomas--Fermi theory gives $\beta^{2} =
3 v_{\mathrm{F}}^{2} /5$, with $v_{\mathrm{F}}$ being the Fermi
velocity, while $\beta^{2} = v_{\mathrm{F}}^{2} /3$ for low
frequencies~\cite{Raza:2015a,Halevi:1995}. The equation of motion is complemented
by the principle of charge conservation; GNOR, in the spirit
of~\cite{Landau-Lifshitz-Pitaevskii}, quantifies $\mathbf{J} (\mathbf{E})$
by extending considerations to include both convective and diffusive
transport of charge, i.e. 
\begin{equation}\label{Eq:cd1}
\frac{\partial}{\partial t} n_{1} + 
\boldsymbol{\nabla} \cdot \left(n_{0} \boldsymbol{\nu} \right) =
D \nabla^{2} n_{1}~,
\end{equation}
where the current density is now given by Fick's law $\mathbf{J} =
(-e) n_{0} \boldsymbol{\nu} - D \boldsymbol{\nabla} (-e) n_{1}$, and $D$
is the diffusion constant. Combining these equations one eventually
obtains
\begin{equation}\label{Eq:NL}
\xi^{2} \boldsymbol{\nabla}
\left( \boldsymbol{\nabla} \cdot \mathbf{J} \right) +
\mathbf{J} =
\sigma_{\mathrm{D}} \mathbf{E}~,
\end{equation}
where $\sigma_{\mathrm{D}} = \mathrm{i} e^{2} n_{0} /
[m (\omega + \mathrm{i} \gamma)]$ is the frequency-dependent Drude
conductivity known from LRA, while
\begin{equation}\label{Eq:NonlocalKsi}
\xi^{2} =
\frac{\beta^{2}}{\omega \left(\omega + \mathrm{i} \gamma \right)} +
\frac{D}{\mathrm{i} \omega} =
\frac{\beta^{2} + D \left(\gamma - \mathrm{i} \omega\right)}
{\omega \left(\omega + \mathrm{i} \gamma\right)}
\end{equation}
is the characteristic nonlocal length. The nonlocal dynamics is now
governed by the coupled equations~(\ref{Eq:Maxwell}) and (\ref{Eq:NL}). 
It is therefore obvious that any such generalised nonlocal model can
be directly implemented in any analytic or numerical formalism developed
for HDM,
or even to effective models that seek to circumvent HDM~\cite{luo_prl111},
just by introducing the generalised complex convection-diffusion
hydrodynamic parameter
\begin{equation}\label{Eq:Eta}
\eta^{2} = \beta^{2} + D \left(\gamma - \mathrm{i} \omega\right)~.
\end{equation}

\section{Boltzmann approach}

In the Boltzmann formalism the induced current density
is given by
\begin{equation}\label{Eq:J}
\mathbf{J} = (-e)
\int \mathrm{d} v \, \boldsymbol{v} f (\boldsymbol{v})~,
\end{equation}
where $\boldsymbol{v}$ is the electron velocity. This is subject
to the continuity equation
\begin{equation}\label{Eq:continuity}
(-e) \frac{\partial}{\partial t} n +
\boldsymbol{\nabla} \cdot \mathbf{J} =
0~.
\end{equation}
Here, $f(\boldsymbol{v})$ is the non-equilibrium distribution
function governed by the Boltzmann equation of motion
\cite{Ashcroft:1976},
\begin{equation}\label{Eq:Boltzmann}
\frac{\partial f}{\partial t} +
\boldsymbol{v} \cdot \boldsymbol{\nabla} f +
\frac{(-e)}{m} \mathbf{E} \cdot
\boldsymbol{\nabla}_{\boldsymbol{v}} f =
I_{\mathrm{col}}[f]~,
\end{equation}
where $I_{\mathrm{col}} [f]$ is the collision functional to be
specified. In the relaxation-time approximation, $I_{\mathrm{col}}
[f] \approx - \gamma (f - f_{0})$ with $f_{0}$ being the equilibrium
distribution function, and we recover the LRA result of Ohm's law, i.e.
$\mathbf{J} \simeq \sigma_{\mathrm{D}} \mathbf{E}$.

In order to proceed beyond LRA, we focus, in the spirit of the
hydrodynamic equation of motion, equation~(\ref{Eq:eom1}), on a gas of
non-interacting electrons subject to electron-impurity collisions
over its entire volume. In this way we derive equations~(\ref{Eq:eom1})
and (\ref{Eq:cd1}) from equation~(\ref{Eq:Boltzmann}). Here, the velocity
field of hydrodynamics is given by the statistically averaged
velocity field ${\boldsymbol \nu} = \left< \boldsymbol{v} \right> =
\int \mathrm{d} \boldsymbol{v} \, \boldsymbol{v} f(\boldsymbol{v})$,
while the density is given by $n = \int \mathrm{d} \boldsymbol{v} \, 
f(\boldsymbol{v})$. 

The procedure to derive equation~(\ref{Eq:eom1}) is to multiply
equation~(\ref{Eq:Boltzmann}) by $\boldsymbol{v}$ and then integrate
over velocity,
\begin{equation}\label{Eq:EOM1}
\frac{\partial}{\partial t} \boldsymbol{\nu} +
\frac{(-e)}{m} \mathbf{E} +
\int \mathrm{d} \boldsymbol{v} \, \boldsymbol{v}
\left( \boldsymbol{v} \cdot \boldsymbol{\nabla} f \right) =
- \gamma \boldsymbol{\nu} ~.
\end{equation}
Assuming that the electron gas is isotropic and that
$\left< v_{j} v_{i} \right> = \left< v_{j}^{2} \right>
\delta_{ji}$, we obtain for the last term on the left-hand side 
\begin{equation}\label{Eq:last-term}
\int \mathrm{d} \boldsymbol{v} \, \boldsymbol{v}
\left(\boldsymbol{v} \cdot \boldsymbol{\nabla} f \right) =
\frac{1}{3} \boldsymbol{\nabla} \left< \boldsymbol{v}^{2} \right> =
\frac{1}{m} \frac{2}{3} \frac{\partial \left< \mathcal{E} \right>}
{\partial n} \boldsymbol{\nabla} n_{1}~,
\end{equation}
where $\left< \mathcal{E} \right> = \left< m \boldsymbol{v}^2 /2 
\right>$ is the average kinetic energy of the free-electron gas. At
zero temperature the kinetic energy of the gas, which arises from
the quantum degeneracy pressure, can be expressed in terms of the
electron density as
\begin{equation}
\left< \mathcal{E}(n) \right> =
\frac{3 \hbar^{2}}{10 m} \left(3\pi^{2} n\right)^{\frac{2}{3}}~.
\end{equation}
This paper only considers the linear dynamics of the Fermi gas.
Linearising equation~(\ref{Eq:last-term}) we obtain
\begin{equation}\label{Eq:linearized}
\int \mathrm{d} \boldsymbol{v} \, \boldsymbol{v}
\left( \boldsymbol{v} \cdot \boldsymbol{\nabla} f \right) =
\frac{\beta^{2}}{n_{0}} \boldsymbol{\nabla} n_{1}~.
\end{equation}

We define the polarisation current as $\mathbf{J}_{\mathrm{e}} =
(-e) \mathbf{E}$. Euation~(\ref{Eq:EOM1}) can be transformed into a
form similar to that of equation~(\ref{Eq:NL}). For a time dependent
electric field, $E(t) = \mathrm{Re} [E(\omega) e^{-\mathrm{i}
\omega t}]$, we can employ a Fourier transformation, alongside the
continuity equation, to obtain
\begin{equation}\label{Eq:DrudeCurrent}
\mathbf{J}_{\mathrm{e}} +
\frac{\beta^{2}}{\omega \left(\omega + \mathrm{i} \gamma \right)}
\boldsymbol{\nabla}
\left( \boldsymbol{\nabla} \cdot \mathbf{J}_{\mathrm{e}} \right) =
\sigma_{\mathrm{D}}
\mathbf{E}~,
\end{equation}
This expression has a Drude form with a nonlocal correction. We note that one, and
only one, nonlocal correction arises from the Boltzmann treatment of the
bulk dynamics. Therefore a diffusion term that is independent of the
quantum pressure convection cannot result from the bulk dynamics of
the system, which is in accordance with the fluctuation-dissipation
theorem~\cite{Kubo:1966}. In the following we consider the high
($\omega \gg \gamma$) and low ($\omega \ll \gamma$) frequency limits
to show that both the diffusive and the convective behaviour are
contained in equation~(\ref{Eq:DrudeCurrent}).

For noble metals the bulk relaxation time is relatively large, implying
a relatively small scattering rate $\gamma$. This means
that when probing the metallic nanostructures at optical frequencies, the
high frequency limit holds true, $\gamma \ll \omega$~\cite{Mortensen:2014}.
Expanding the prefactor of the second term on the left-hand side of
equation~(\ref{Eq:DrudeCurrent}) (which gives a measure of the strength of
nonlocality) to first order in $\gamma$ results in
\begin{equation}\label{Eq:NonlocalityExpansionHigh}
\frac{\beta^{2}}{\omega \left(\omega + \mathrm{i} \gamma\right)}
\approx \frac{\beta^{2}}{\omega^{2}} -
\mathrm{i} \frac{\gamma \beta^{2}}{\omega^{3}} +
\mathcal{O}(\gamma^{2})~.
\end{equation}
Because the imaginary part on the right-hand side of
equation~(\ref{Eq:NonlocalityExpansionHigh}) (which describes diffusion)
scales as $\sim \omega^{-3}$, the damping due to nonlocality arising
from the bulk properties of the material vanishes in the high frequency
limit. The nonlocal behaviour in the high (optical) frequency limit is
therefore purely convective.
 
On the other hand, in the low frequency limit terms of order
$\mathcal{O}(\omega^{2})$ can be neglected and one obtains
\begin{equation}\label{Eq:NonlocalityExpansionLow}
\frac{\beta^{2}}{\omega \left(\omega + \mathrm{i} \gamma \right)}
\approx \frac{\beta^{2}}{\mathrm{i} \omega \gamma}~.
\end{equation}
The above expression has the same form as the diffusive nonlocal
correction in equation~(\ref{Eq:NonlocalKsi}), leading to a diffusion
constant of the form
\begin{equation}\label{Eq:DiffusionBoltzmann1}
D \equiv \frac{\beta^{2}}{\gamma}~.
\end{equation}
The following equation of motion is then obtained
\begin{equation}\label{Eq:DiffusiveDrude}
\mathbf{J}_{\mathrm{e}} +
\frac{D}{\mathrm{i} \omega} \boldsymbol{\nabla}
\left(\boldsymbol{\nabla} \cdot \mathbf{J}_{\mathrm{e}} \right) =
\sigma_{\mathrm{D}}
\mathbf{E}~.
\end{equation}
Substituting $\gamma = v_\mathrm{F}/\ell$, where $\ell$ is the
electron mean-free path, we end up with
\begin{equation}\label{Eq:DiffusionBoltzmann2}
D = \frac{3}{5} v_\mathrm{F} \ell~.
\end{equation}
This expression for $D$ is consistent with the one anticipated
in~\cite{Mortensen:2014}, namely $D \propto v_{\mathrm{F}} \ell$.
Importantly, however, we are also led to the conclusion that, at
high, optical frequencies, the diffusive term cannot arise
independently from the bulk properties of the plasmonic nanostructures.
It is therefore reasonable to assume that, even though diffusion in
GNOR is introduced in the description of the bulk metal, it will
only be important at the surface, and it is related to surface damping
effects. The importance of interface and local porosity effects has
been recently discussed through a combination of optical and electronic
spectroscopies~\cite{Campos:2019}.

\section{Surface scattering}\label{Sec:Surf}

\begin{figure}[ht]
\centering
\includegraphics[width = 0.48\textwidth]{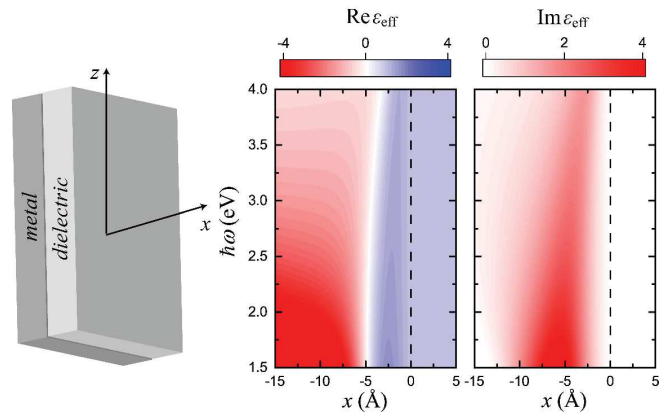}
\caption{Schematic drawing of a metal-dielectric interface (left-hand panel).
Real (middle panel) and imaginary part (right-hand panel) of the effective
dielectric function $\varepsilon_{\mathrm{eff}}$ (its component normal to
the interface) calculated with GNOR for a Drude metal with $\hbar
\omega_{\mathrm{p}} =5$\,eV, $\hbar \gamma =0.025$\,eV, $v_{\mathrm{F}} =
1.07 \cdot 10^{6}$\,m\,s$^{-1}$, and $D = 2 \cdot 10^{-4}$\,m$^{2}$\,s$^{-1}$.
The dielectric is assumed to be air, described by a dielectric constant equal
to one.}\label{Fig1}
\end{figure}

Near the metal surface translational invariance in the normal
direction is broken and momentum conservation no longer applies. The
induced polarisation of the metal can therefore contain components of
finite wavenumber $q$, even if such components are not present in the
external field. From the random-phase approximation (RPA) it is known
that the new field components at $q \geq \omega/v_{\mathrm{F}} \equiv q_{0}$
will be Landau-damped~\cite{Rammer:1986}. The effect of the surface is
therefore to introduce enhanced damping of the plasmon modes, thus
broadening the resonances. This is in agreement with the experimental
findings of Kreibig \emph{et al.}~\cite{Kreibig:1985}, who measured
a $1/R$-dependent broadening of the plasmon resonances for small
metallic nanospheres of radius $R$. This behaviour is commonly
considered within the Kreibig SDB model through the simple
substitution
\begin{equation}\label{Eq:SDB}
\gamma \rightarrow \gamma + v_{\mathrm{F}} \frac{A}{R}~,
\end{equation}
where $A$ is a dimensionless constant, typically taken between 0.5 and 1. This
kind of broadening is reproduced by the GNOR theory through
introduction of the diffusion term. More specifically, it is found
that for a spherical particle, the relaxation time correction to
first order in $1/R$, $\gamma^{(1)}$, is~\cite{Mortensen:2014}
\begin{equation}\label{Eq:gamma1GNOR}
\gamma^{(1)} =
\frac{\sqrt{6}}{24} \frac{D \omega_{\mathrm{p}}}{\beta R}~,
\end{equation}
where $\omega_{\mathrm{p}}$ is the plasma frequency of the
metal. As we saw in the previous section, the diffusion term
cannot arise independently from the bulk properties at optical
frequencies. It is therefore reasonable to postulate that the
GNOR diffusion term must arise by the surface-enhanced Landau
damping, described by the equation of motion
\begin{equation}\label{Eq:GNORmotion}
\mathbf{J}_{\mathrm{e}} +
\left[\frac{\beta^{2}}{\omega (\omega + \mathrm{i} \gamma)} +
\frac{D_{\mathrm{surf}}} {\mathrm{i} \omega} \right]
\boldsymbol{\nabla}
\left( \boldsymbol{\nabla} \cdot \mathbf{J}_{\mathrm{e}} \right) =
\sigma_{\mathrm{D}}
\mathbf{E}~.
\end{equation}
Here, strictly speaking, $D_{\mathrm{surf}}$ is a surface diffusion
constant. Since, however, the hydrodynamic description is extended
in the bulk of the metal, we want to write $D_{\mathrm{surf}} = D$.
This assumption is consistent with the results of Fig.~\ref{Fig1},
where, for the metal-air interface shown in the schematics, we
calculate with GNOR, following the examples by Wubs and Mortensen
in~\cite{Wubs:2016} and Tserkezis \emph{et al.}
in~\cite{Tserkezis:2017b}, an effective dielectric function
$\varepsilon_{\mathrm{eff}} (\omega)$ (we focus on the component
normal to the interface here), assuming for the metal a plasmon
energy $\hbar \omega_{\mathrm{p}} = 5$\,eV, $\hbar \gamma = 0.025$\,eV,
$v_{\mathrm{F}} = 1.07 \cdot 10^{6}$\,m\,s$^{-1}$, and a typical value
for the diffusion constant $D = 2 \cdot 10^{-4}$\,m$^{2}$\,s$^{-1}$.
These values are chosen as typical for good jellium metals, and do not
represent any specific material. More precise values to describe Na will
be used later on. From Fig.~\ref{Fig1} it is evident that the imaginary
part of $\varepsilon_{\mathrm{eff}}$ is large mostly near the interface,
in agreement with the picture of surface-enhanced damping.

The question that still lingers is what expression one should use
for $D$. The value $D \propto \beta^{2}/ \gamma$ [see
equation~(\ref{Eq:NonlocalityExpansionLow})] was proposed in the original
GNOR paper~\cite{Mortensen:2014}, while an approach based on direct
comparison with the broadening predicted by the Kreibig correction
was adopted in \cite{Raza:2015b}. As we showed earlier,
equation~(\ref{Eq:NonlocalityExpansionLow}) is consistent
with what is expected for diffusive electron transport. However it has
yet to be justified in terms of surface dynamics, and must therefore be
examined in more detail. In the following we show that the diffusion
constant can be related to the dissipative part of the induced surface
charge via the Feibelman parameter $d_{\perp}$~\cite{Feibelman:1982}. A
similar connection to the Feibelman parameters has been presented for the
estimation of the resonance energy in~\cite{Raza:2015b}.

Before resorting to the centroid of the induced charge, as introduced
by the Feibelman formalism, it is intuitive to consider a simpler
approach based on the general, wavevector $\mathbf{q}$-dependent Lindhard
dielectric function~\cite{Lindhard:1954}
\begin{equation}\label{Eq:Lindhard}
\varepsilon (\omega, \mathbf{q}) =
\varepsilon_{\mathrm{b}} +
\frac{3 \omega_{\mathrm{p}}^{2}}{q^{2} v_{\mathrm{F}}^{2}}
\left[1 - \frac{\omega}{2q v_{\mathrm{F}}}
\ln \frac{\omega + q v_{\mathrm{F}}}{\omega - q v_{\mathrm{F}}}
\right]~.
\end{equation}
We revisit the procedure described by Sun and Khurgin
in~\cite{Bozhevolnyi:2016a}, and assume that surface scattering
is much more important than collisions in the bulk and the
relaxation rate $\gamma$ can be taken entirely due to surface effects.
Because Landau damping only affects the field components with $q > q_{0}$,
an approximation is needed to capture the effect with an effective
relaxation time. From the imaginary part of the Drude expression
one can relate the relaxation time to the imaginary part of the
dielectric function, $\mathrm{Im} \, \varepsilon$, to obtain
\begin{equation}\label{Eq:GammaKhurgin}
\gamma =
\frac{\omega^{3} \mathrm{Im} \, \varepsilon}
{\omega_{\mathrm{p}}^{2}}~.
\end{equation}
The longitudinal components of the electric field with
$q > q_{0}$ will experience Landau damping. To eliminate the
$q$-dependence of the Lindhard dielectric function, an effective
dielectric function is now defined through
\begin{equation}\label{Eq:EpsilonEffective}
\varepsilon_{\mathrm{eff}} (\omega) =
\frac{\int_{q_{0}}^{\infty}
\mathrm{d}^{3} q \,
\varepsilon (\omega, \mathbf{q}) 
\left|E_{\parallel} (\mathbf{q}, \omega)\right|^{2}}
{\int_{0}^{\infty} \mathrm{d}^{3} q \,
\left|E(\mathbf{q},\omega)\right|^{2}}~.
\end{equation}
The Lindhard dielectric function has a finite imaginary part
for $q > q_{0}$ given as
\begin{equation}\label{Eq:EpsilonImag}
\mathrm{Im}\,\varepsilon =
\frac{3 \omega_{\mathrm{p}}^{2}}{2\omega^{2} q^{3}}~.
\end{equation}
Assuming again the simple metal-dielectric interface of
Fig.~\ref{Fig1}, a propagating surface plasmon is considered,
whose electric field inside the metal is given by $E(x, z) = E_{0}
\exp(-x/d_{\mathrm{s}}) \exp(\mathrm{i}kz)$, where $d_{\mathrm{s}}$
is the skin depth of the metal. For a typical value of $v_{\mathrm{F}}
\sim 10^{6}$\,m\,s$^{-1}$ and $d_{\mathrm{s}} \sim 10^{-8}$\,m, we
then obtain the following effective relaxation rate due to the surface
damping
\begin{equation}\label{Eq:GammaKhurgin2}
\gamma =
\frac{3 v_{\mathrm{F}}}{4 d_{\mathrm{s}}}
\sim 10^{14}\,
\mathrm{s}^{-1} \;\;
\Leftrightarrow \;\;
\hbar \gamma \sim 0.414\,\mathrm{eV}~.
\end{equation}
In nanostructures where one of the characteristic lengths is
smaller that the skin depth of the metal, $d_{\mathrm{s}}$ should
of course be replaced by this length.

\begin{figure}[ht]
\centering
\includegraphics[width = 0.48\textwidth]{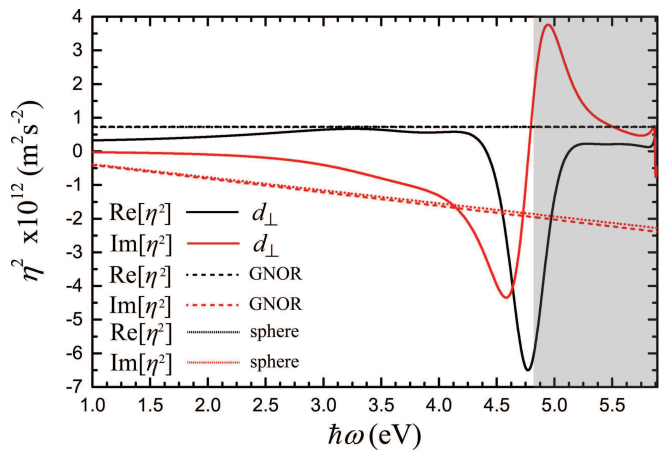}
\caption{Complex convective-diffusive hydrodynamic parameter
$\eta^{2}$ (real part: black lines; imaginary part: red lines), as
obtained by the $d_{\perp}$ method [equation~(\ref{Eq:dhydro})] from the
reflection coefficient of a flat air-Na interface (solid lines), from
the GNOR diffusion constant obtained from $d_{\perp}$ for a sphere
through equation~(\ref{Eq:DFeibelman}) (dotted lines), or for a constant
diffusion parameter $D = 2.67 \cdot 10^{-4}$ \,m$^{2}$\,s$^{-1}$~\cite{Raza:2015a}
(dashed lines). In all cases, we have assumed a
jellium metal with $r_{\mathrm{s}} =4$ and $\hbar \omega_{\mathrm{p}} =$
5.89143\;eV, $\hbar \gamma =$ 0.1\;eV, $v_{\mathrm{F}} = 1.07 \cdot
10^{6}$\,m\,s$^{-1}$). The Feibelman $d_{\perp}$ parameter is obtained
from~\cite{Christensen:2017}. The shaded area denotes the frequency
range (around the Bennett resonance, and close to the plasma frequency)
where a calculation of $\eta^{2}$ based on equation~(\ref{Eq:dhydro})
is ambiguous  and cannot be trusted.
}\label{Fig2}
\end{figure}

Let us now turn to the more microscopic Feibelman approach~\cite{Feibelman:1982,Christensen:2017,Goncalves:2020},
and relate the induced surface charge to the resonance broadening.
The Feibelman $d$ parameters have been recently proven an efficient route
towards quantum plasmonics~\cite{Yang:2019,Goncalves:2020}, as they are
capable of capturing all screening, Landau damping, and spill-out, which
are the  dominant effects any quantum-informed model should be able
to address~\cite{Tserkezis:2018}. Christensen \emph{et al.}~\cite{Christensen:2017,Goncalves:2020}
have shown that the first-order correction to the damping rate can be connected
to the imaginary part of the perpendicular $d$ parameter, $d_{\perp}$, which
physically represents the centroid of induced charge, through
\begin{equation}\label{Eq:Gamma1Feibelman}
\gamma^{(1)} =
-\frac{1}{4} \frac{\omega_{\mathrm{p}}^{2}}{\omega^{(0)}}
\Lambda_{\perp} \mathrm{Im} \, d_{\perp}^{(0)}~,
\end{equation}
where $\Lambda_{\perp}$ is a geometric parameter, and the superscript
${(0)}$ indicates that the corresponding function is to be calculated
on resonance. For the dipole resonance in a sphere, $\omega^{(0)} \simeq
\omega_{\mathrm{p}}/\sqrt{3}$, and the correction to the damping rate
becomes
\begin{equation}\label{Gamma1Feibelman2}
\gamma^{(1)} = \frac{1}{3 R}\frac{\omega_{\mathrm{p}}^{2}}{\omega^{(0)}}
\mathrm{Im} \, d_{\perp}^{(0)}~.
\end{equation}
Comparing with the GNOR expression of equation~(\ref{Eq:gamma1GNOR}) one
can obtain a first approximate connection between the diffusion constant
and the induced surface charge:
\begin{equation}\label{Eq:DFeibelman}
D = \frac{8 \beta }
{\sqrt{2} } \mathrm{Im} \, d_{\perp}^{(0)}. 
\end{equation}
In Fig.~\ref{Fig2} we plot with dotted lines the complex hydrodynamic
parameter $\eta^{2}$ obtained based on equation~(\ref{Eq:Eta}), for the
diffusion constant derived from equation~(\ref{Eq:DFeibelman}), using
the Feibelman parameter derived in~\cite{Christensen:2017} for a
jellium metal with Wigner--Seitz radius $r_{\mathrm{s}} = 4$, which
well represents Na. The plasma frequency corresponding to this radius
is $\hbar \omega_{\mathrm{p}} = 5.89143$ \; eV, and the Fermi velocity
is $v_{\mathrm{F}} = 1.07 \cdot 10^{6}$\,m\,s$^{-1}$, while for the
Drude damping rate we assume $\hbar \gamma = 0.1$ \; eV.
We note here that we focus on Na, even though hydrodynamics is known
to fail to predict the correct frequency shifts in its case -- as will
be shown in Fig.~\ref{Fig3} -- because Feibelman
parameters have only been unambiguously derived for jellium metals.
Obtaining such parameters for noble metals is a more challenging
task, and constitutes part of our ongoing activities.

The $D$ obtained from equation~(\ref{Eq:DFeibelman}) can be a good starting
point for using the GNOR model. Most notably, it turns out that it
is in very good agreement with the corresponding results obtained
assuming a constant diffusion constant $D$ chosen to match the damping
of the SDB model~\cite{Raza:2015a} (dashed lines in Fig.~\ref{Fig2}).
Nevertheless, it is still derived based on just the dipolar mode of a
spherical nanoparticle, and extending its use to different shapes or
even higher-order modes of spheres is not straightforward to justify,
as it assumes that the damping is the same in all cases and all frequencies.
A more general, frequency-dependent approach, which only disregards curvature
effects, could be based on the reflection coefficients of a flat
metal-dielectric interface. In the $d$-parameter formalism, the reflection
coefficient for $p$ polarisation is given by~\cite{Feibelman:1982,Goncalves:2020}
\begin{eqnarray}\label{Eq:rFeibelman}
&&r_{d_{\perp}, p} =\nonumber\\
&& \frac{\varepsilon_{\mathrm{m}} k_{x,\mathrm{d}} -
\varepsilon_{\mathrm{d}} k_{x,\mathrm{m}} +
\left( \varepsilon_{\mathrm{m}} - \varepsilon_{\mathrm{d}}\right)
\left[\mathrm{i} q^{2} d_{\perp} -
\mathrm{i} k_{x, \mathrm{d}} k_{x, \mathrm{m}} d_{\parallel} \right]}
{\varepsilon_{\mathrm{m}} k_{x,\mathrm{d}} +
\varepsilon_{\mathrm{d}} k_{x,\mathrm{m}} -
\left( \varepsilon_{\mathrm{m}} 
- \varepsilon_{\mathrm{d}}\right) \left[\mathrm{i} q^{2} d_{\perp} +
\mathrm{i} k_{x, \mathrm{d}} k_{x, \mathrm{m}} d_{\parallel} \right]}, \nonumber\\
\end{eqnarray}
where $d_{\parallel}$ is the parallel $d$ parameter (centroid of induced
current), $q$ is now the in-plane wavenumber and $k_{j} = \sqrt{\varepsilon_{j}}
k_{0}$ is the bulk wavenumber, with $j = \mathrm{d}, \mathrm{m}$ denoting
the medium, dielectric or metal, and $k_{0} = \omega/c$ is the free-space
wavenumber; $\varepsilon_{j}$ is the (complex and dispersive, but local) 
permittivity of medium $j$, and $k_{x,j} = \sqrt{k_{j}^{2} - q^{2}}$ is
the wavenumber in the out-of-plane direction, taken to be the $x$ direction
(see Fig.~\ref{Fig1}). On the other hand, in a generalised hydrodynamic
formalism, the corresponding reflection coefficient reads~\cite{David:2013}
\begin{eqnarray}\label{Eq:rHydro}
r_{h, p} = \frac{\varepsilon_{\mathrm{m}} k_{x,\mathrm{d}} -
\varepsilon_{\mathrm{d}} k_{x,\mathrm{m}} \left(1+ \delta_{h} \right)}
{\varepsilon_{\mathrm{m}} k_{x,\mathrm{d}} +
\varepsilon_{\mathrm{d}} k_{x,\mathrm{m}} \left(1 + \delta_{h} \right)}
~,
\end{eqnarray}
with $\delta_{h} = [q^{2} (\varepsilon_{\infty} -
\varepsilon_{\mathrm{m}})]/(k_{x,\mathrm{m}} k_{x,\mathrm{nl}}
\varepsilon_{\infty})$ and $k_{x,\mathrm{nl}}^{2} =
k_{\mathrm{nl}}^{2} - q^{2} = [\omega (\omega + \mathrm{i} \gamma) -
\omega_{\mathrm{p}}^{2}/ \varepsilon_{\infty}]/\eta^{2}  - q^{2} $. In
all cases, the imaginary part of $k_{x}$ must be larger than zero, to
ensure a passive medium. Direct comparison between Eqs.~(\ref{Eq:rFeibelman})
and (\ref{Eq:rHydro}) in the long-wavelength limit ($q \rightarrow 0$)
yields~\cite{PrivateCommun}
\begin{equation}\label{Eq:dhydro}
d_{\perp} = - \frac{\mathrm{i}}{k_{\mathrm{nl}}}
\frac{\varepsilon_{\mathrm{d}}}{\varepsilon_{\infty}}
\frac{\varepsilon_{\mathrm{m}} - \varepsilon_{\infty}}
{\varepsilon_{\mathrm{m}} - \varepsilon_{\mathrm{d}}}~.
\end{equation}
Solving for $\eta^{2}$, with the constraints that $\mathrm{Im}\,k_{\mathrm{nl}}$
should be positive~\cite{Feibelman:1982} and
the imaginary part of  $\eta^{2}$, as introduced in
GNOR~\cite{Mortensen:2014} needs to be negative to ensure a
lossy medium, one can obtain a quantum-informed, generalised
dispersive expression for the complex convective-diffusive
hydrodynamic parameter, which can apply equally well to all
shapes of
nanostructures~\cite{wiener_nl12}.
In Fig.~\ref{Fig2} this $\eta^{2}$
is plotted as a function of frequency with solid lines.

Up to around 4.7\;eV the results obtained through equation~(\ref{Eq:dhydro})
are close to those corresponding to constant diffusion parameters,
predicting a slightly lower damping in the low-frequency limit. This
behaviour is maintained as the frequency increases, and one approaches
the region of interest, where all plasmonic resonances are expected
(the dipolar resonance of a Na sphere in the quasistatic limit is
around 3.4\;eV). For even higher frequencies, however, the $d$
parameter-based $\eta^{2}$ becomes more ambiguous, especially around
the so-called Bennett multipole surface plasmon~\cite{Toscano:2015,Bennett:1970}
which appears due to spill-out at 4.7\;eV. In this region, and all the
way up to the plasma frequency (grey-shaded area in Fig.~\ref{Fig2}),
the requirement that the imaginary part of the retrieved $\eta^{2}$
be negative leads to both real and imaginary part appearing with the
opposite sign of what is shown in the figure. In that case, instead
of the typical shape of a Lorentzian around a resonance, one would
get an unnatural, kinked curve, which obviously violates
the Kramers--Kronig relations~\cite{Ashcroft:1976}. So one of the
two requirements has to be relaxed, and since Kramers--Kronig relations
are a manifestation of causality, we choose to comply with them.
Doing this, the retrieval process is forced to introduce a fictitious
gain,
in agreement with the local equivalent-layer approach of
\cite{luo_prl111}
[note that when co-existing with the even more lossy response of the
bulk, it still results in a net lossy response, see equation~(\ref{Eq:NonlocalKsi})].
Obviously, asking a hydrodynamic description of the bulk, with
spill-in intrinsically built into it, to reproduce all the physics
captured by a surface-response function which predicts spill-out
has its limitations, and the results in the grey-shaded area cannot
be trusted. Our approach has, nevertheless, provided a link between
other phenomenological approaches to quantify surface-enhanced
Landau damping, and they all appear consistent up to
$\simeq 0.8 \omega_{\mathrm{p}}$.

\section{Numerical results}\label{Sec:Appl}

\begin{figure}[ht]
\centering
\includegraphics[width = 0.48\textwidth]{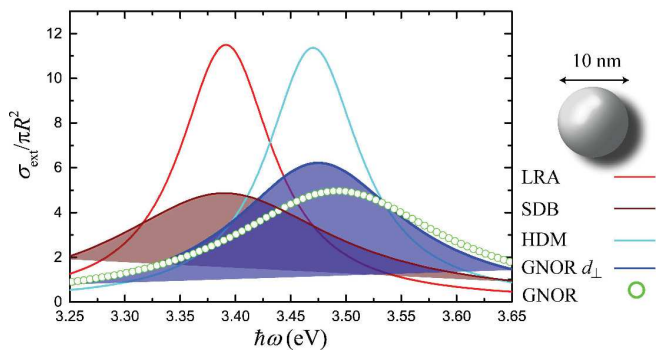}
\caption{Normalised extinction cross section for a Na sphere with
radius $R = 5$\;nm, as obtained within LRA (red line), SDB (dark-red
shaded line), HDM (light blue line), GNOR with a $d_{\perp}$-based
hydrodynamic parameter (dark-blue shaded line) and GNOR with a constant
diffusion $D = 2.67 \cdot 10^{-4}$ \,m$^{2}$\,s$^{-1}$ (open green
circles).}\label{Fig3}
\end{figure}

To illustrate the predictions of our analysis and the value
of the $d$ parameter-based calculation of $\eta^{2}$, we present
in Fig.~\ref{Fig3} 
analytic calculations of
normalised extinction cross section spectra for
a Na sphere described by a Drude model with the parameters mentioned
at the end of Sec.~\ref{Sec:Surf}. The radius of the sphere is taken
equal to $R = 5$\;nm, and the spectra are calculated with an
implementation of GNOR based on Mie theory~\cite{Tserkezis:2016a},
using either the hydrodynamic parameter obtained by equation~(\ref{Eq:Eta})
(dark-blue shaded line) or the constant diffusion parameter $D$
suggested in~\cite{Raza:2015a} (green open circles). The results are
compared to those of LRA (red line), together with the standard
HDM (light blue line) and the SDB phenomenological correction
(dark-red shaded line, with $A = 1.0$). The spectrum obtained by
calculating $\eta^{2}$ using the Feibelman-based recipe outlined
above agrees very well in resonant frequency with the prediction
of HDM, and gives slightly smaller broadening as compared to SDB
(whose constant $A$ contains a certain arbitrariness anyway). The
open green circles already show that the effect of the surface-enhanced
Landau damping can in principle be captured using a \emph{constant}
GNOR diffusion term, but the full connection to the centroid of induced charge
can be more accurate, as it contains the quantum-mechanically
calculated frequency dependence of the hydrodynamic parameters. 

\begin{figure}[ht]
\centering
\includegraphics[width = 0.48\textwidth]{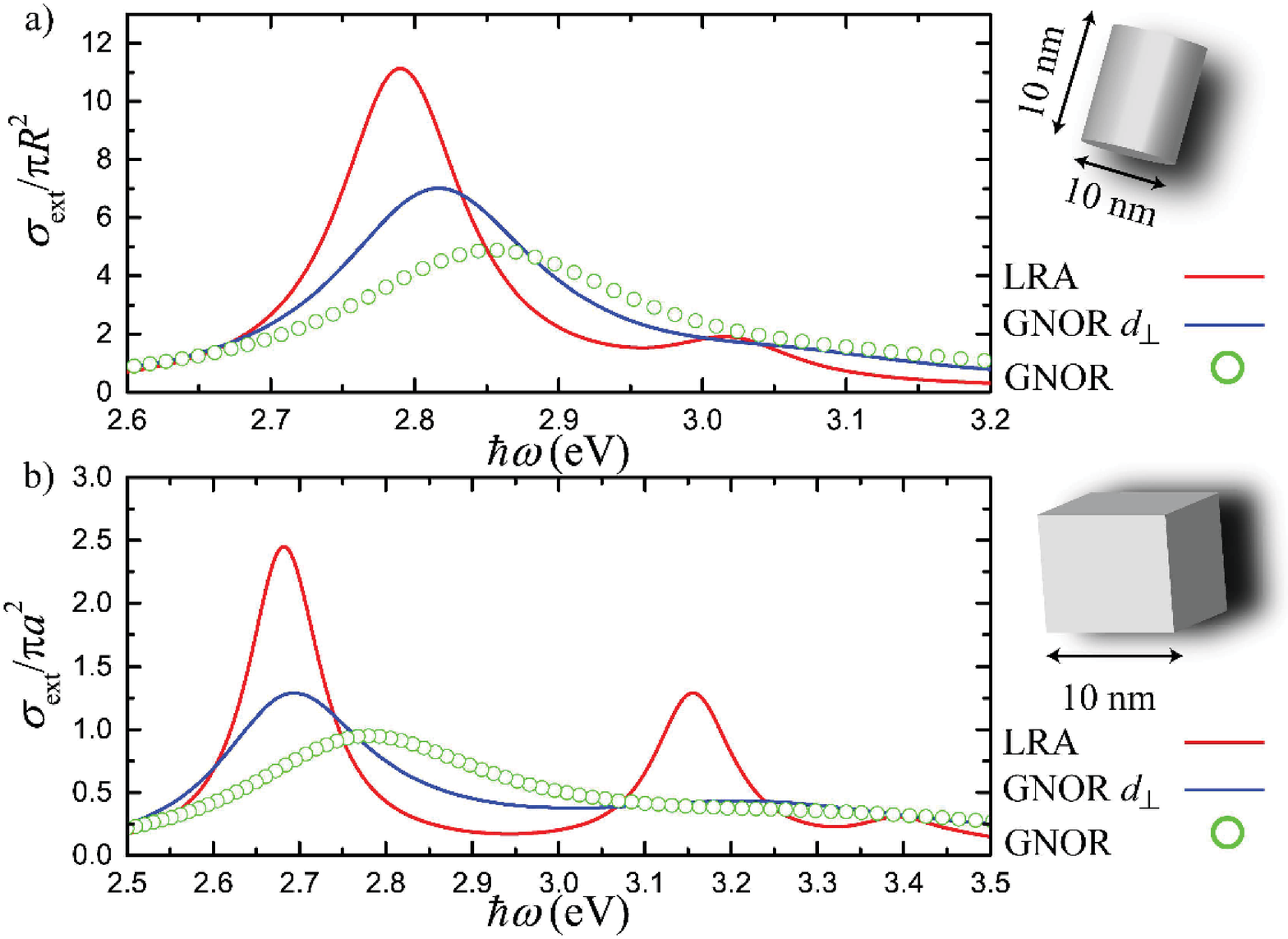}
\caption{(a) Normalised extinction cross section spectra for a
cylinder with diameter $R/2 =10$\;nm and height $H = 10$\;nm,
as obtained within LRA (red line), GNOR with a $d_{\perp}$-based
hydrodynamic parameter (dark blue line), and GNOR with constant
diffusion $D = 2.67 \cdot 10^{-4}$ \,m$^{2}$\,s$^{-1}$ (green
open circles). The incident field is polarised at an angle $\theta =
30^{\circ}$ with respect to the cylinder axis. (b) Same as (a), for
a cube with side $a = 10$\;nm. The incident field is polarised along
one of the cube sides.}\label{Fig4}
\end{figure}

As further demonstration of the potential of our derivation, we
study in Fig.~\ref{Fig4} two nanoparticles of different shapes:
a cylinder with diameter $R/2$ and height $H$ both equal to 10\;nm
(a), and a cube with side $a =$ 10\;nm (b). The spectra are obtained
using our implementation of hydrodynamic methods to a commercial
finite-element solver (Comsol Multiphysics 5.1), with the only change
compared to previous studies~\cite{Tserkezis:2018,Tserkezis:2016a,
Tserkezis:2017} being the calculation of $\eta^{2}$, which is now
performed not based on Eq.~(\ref{Eq:Eta}) for constant $\beta$ and
$D$, but by solving Eq.~(\ref{Eq:dhydro}) for the $d_{\perp}$ values
obtained from~\cite{Christensen:2017} (as interpolated by an analytic
function in~\cite{Goncalves:2020}).
The NPs were modelled with approximately $100000$ tetrahedral domain elements
with minimum side of $0.2$\;nm, while perfectly-matched layers of thickness 
$300$\;nm were used to minimise reflections at the simulation area boundaries.
The results are again compared
both to the LRA approximation and the traditional GNOR approach with
constant diffusion. In both cases, it is evident that the constant-$D$
results overestimate both the damping and the blueshift of the modes,
a response which could be anticipated by the differences between the
corresponding $\eta^{2}$ values in Fig.~\ref{Fig2}, especially in the
region of higher order (edge or corner) modes~\cite{Grillet:2011},
which are always affected more by damping.

The above analysis and results clearly display the main strength of
the Feibelman-based approach: because $d_{\perp}$ is obtained for
a planar interface, one quantum-mechanical (e.g. with density-functional theory)
calculation per material should be sufficient
to obtain the hydrodynamic parameter $\eta^{2}$, which can subsequently
be introduced to hydrodynamic calculations in arbitrary geometries,
as already implemented in a series of different
numerical~\cite{McMahon:2010a,Toscano:2012,Trugler:2017,Li:2017,Doicu:2020}
or approximate methods~\cite{luo_prl111}.
In addition to being easily generalisable, the results presented
in this article also establish a solid link between the surface-enhanced
Landau damping and the electron convection-diffusion mechanisms, thus
providing a theoretical justification of the introduction of the
GNOR diffusion term.

\section{Conclusion}

We have discussed the origin and role of the diffusion term included
in the GNOR theory for nonlocal plasmonics. Based on the Boltzmann
equation, we showed that the diffusion-induced damping in GNOR cannot
have its origins in the bulk of the material, but should be considered
as an efficient way to capture surface-enhanced Landau damping. We
linked the complex hydrodynamic convection-diffusion parameter to
the induced surface charge via the Feibelman $d_{\perp}$ parameter
for the centroid of induced charge, enabling the extraction of the input
required for hydrodynamic models from systematic \textit{ab-initio}
calculations. The results obtained for a single metal-dielectric
interface can be implemented in any geometry, thus allowing to
expand the use of GNOR in a wide range of applications in
nonclassical plasmonics.

\section*{Acknowledgements}
We thank T.~Christensen and P.~A.~D.~Gon\c{c}alves for valuable discussions,
and G.~W.~ Hanson for early important discussions that stimulated our work based on the Boltzmann method.
NAM is a VILLUM Investigator supported by VILLUM FONDEN (grant No. 16498).
The Center for Nano Optics is financially supported by the University
of Southern Denmark (SDU 2020 funding).
The Center for Nanostructured Graphene is sponsored by the Danish
National Research Foundation (Project No. DNRF103).
CT is indebted to W. Berry, P. Buck, M. Mills and M. Stipe for
lifelong inspiration.
CW acknowledges funding from MULTIPLY fellowships under the Marie
Skłodowska-Curie COFUND Action (grant agreement No. 713694).
Simulations were supported by the DeIC National HPC Centre, SDU.


\section*{References}
\begin{thebibliography}{99}
\bibitem{Tame:2013} Tame M~S, McEnery K~R, \"{O}zdemir \c{S}~K, Lee J,
Maier S~A and Kim M~S
\textit{Nature Phys.} 2013 \textbf{9} 329--340 
%
\bibitem{Bozhevolnyi:2016a} Bozhevolnyi S~I, Mart\'{i}n-Moreno L and
Garc\'{i}a-Vidal F~J 2017
\textit{Quantum Plasmonics} (Springer)
%
\bibitem{Bozhevolnyi:2016b} Bozhevolnyi S~I and Mortensen N~A
\textit{Nanophotonics} 2017 \textbf{6} 1185--1188
%
\bibitem{Zhu:2016} Zhu W, Esteban R, Borisov A~G, Baumberg J~J,
Nordlander P, Lezec H~J, Aizpurua J and Crozier K~B
\textit{Nature Commun.} 2016 \textbf{7} 11495
%
\bibitem{Fitzgerald:2016} Fitzgerald J~M, Narang P, Craster R~V,
Maier S~A and Giannini V
\textit{Proc. IEEE} 2016 \textbf{104} 2307--2322
%
\bibitem{Yannopapas:2016} Yannopapas V
\textit{Int. J. Mod. Phys. B} 2017, \textbf{31} 1740001
%
\bibitem{Amendola:2017} Amendola V, Pilot R, Frasconi M,
Marag\`{o} O~M and Iat\`{i} M~A
\textit{J. Phys.: Condens. Matter} 2017 \textbf{29} 203002
%
\bibitem{Fernandez-Dominguez:2018} Fern{\'a}ndez-Dom{\'i}nguez A I,
Bozhevolnyi S I and Mortensen N A
\textit{ACS Photonics} 2018 \textbf{5} 3447
%
\bibitem{Ciraci:2019} Cirac\`{i} C, Jurga R, Khalid M and Della Sala F
\textit{Nanophotonics} 2019 \textbf{6} 1821--1833
%
\bibitem{Savage:2012} Savage K~J, Hawkeye M~M, Esteban R, Borisov A~G,
Aizpurua J and Baumberg J~J
\textit{Nature} 2012, \textbf{491} 574--577
%
\bibitem{Duan:2012} Duan H, Fern\'{a}ndez-Dom\'{i}nguez A~I, Bosman M,
Maier S~A and Yang J~K~W
\textit{Nano Lett.} 2012, \textbf{12} 1683--1689
%
\bibitem{Scholl:2012} Scholl J~A, Koh A~L and Dionne J~A
\textit{Nature} 2012, \textbf{483} 421--427
%
\bibitem{Scholl:2013} Scholl J~A, Garc\'{i}a-Etxarri A,
Koh A~L and Dionne J~A
\textit{Nano Lett.} 2013 \textbf{13} 564--569
%
\bibitem{Wiener:2013} Wiener A, Duan H, Bosman M, Horsfield A~P,
Pendry J~B, Yang J~K~W, Maier S~A and Fern\'{a}ndez-Dom\'{i}nguez A~I
\textit{ACS Nano} 2013 \textbf{7} 6287--6296
%
\bibitem{Jung:2015} Jung H, Cha H, Lee D and Yoon S
\textit{ACS Nano} 2015 \textbf{9} 12292--12300
%
\bibitem{Raza:2015c} Raza S, Kadkhodazadeh S, Christensen T,
Di Vece M, Wubs M, Mortensen N~A and Stenger N,
\textit{Nature Commun.} 2015 \textbf{6} 8788
%
\bibitem{Krasavin:2016} Krasavin A~V, Ginzburg P, Wurtz G~A
and Zayats A~V
\textit{Nature Commun.} 2016 \textbf{7} 11497
%
\bibitem{Shen:2017} Shen H, Chen L, Ferrari L, Lin M-H,
Mortensen N~A, Gwo S and Liu Z,
\textit{Nano Lett.} 2017 \textbf{17} 2234--2239
%
\bibitem{Abajo:2008} Garc\'{i}a de Abajo F~J
\textit{J. Phys. Chem. C} 2008\textbf{112} 17983--17987
%
\bibitem{McMahon:2010b} McMahon J~M, Gray S~K and Schatz G~C
\textit{Nano Lett.} 2010 \textbf{10} 3473--3481
%
\bibitem{Raza:2011} Raza S, Toscano G, Jauho A-P,
Wubs M and Mortensen N~A
\textit{Phys. Rev. B} 2011 \textbf{84} 121412(R)
%
\bibitem{Ciraci:2012} Cirac\`{i} C, Hill R~T, Mock J~J, Urzhumov Y,
Fern\'{a}ndez-Dom\'{i}nguez A~I, Maier S~A, Pendry J~B, Chilkoti A
and Smith D~R
\textit{Science} 2012 \textbf{337} 1072--1074
%
\bibitem{David:2014} David C and Garc\'{i}a de Abajo F~J
\textit{ACS Nano} 2014 \textbf{8} 9558--9566
%
\bibitem{Christensen:2014} Christensen T, Yan W, Raza S, Jauho A-P,
Mortensen N~A and Wubs M
\textit{ACS Nano} 2014 \textbf{8} 1745--1758
%
\bibitem{Raza:2015a} Raza S, Bozhevolnyi S~I, Wubs M and Mortensen N~A
\textit{J. Phys.: Condens. Matter} 2015 \textbf{27} 183204
%
\bibitem{Eremin:2018} Eremin Y, Doicu A and Wriedt T
\textit{J. Quant. Spectr. Rad. Transf.} 2018 \textbf{217} 35--44
%
\bibitem{Kupresak:2020} Kupresak M, Zheng X, Vandenbosch G~A~E
andMoshchalkov V~V
\textit{Adv. Theory Simul.} 2020 \textbf{3} 1900172
%
\bibitem{Esteban:2012} Esteban R, Borisov A~G, Nordlander P and
Aizpurua J
\textit{Nature Commun.} 2012 \textbf{3} 825
%
\bibitem{Stella:2013} Stella L, Zhang P, Garc\'{i}a-Vidal F~J,
Rubio A and Garc\'{i}a-Gonz\'{a}lez P
\textit{J. Phys. Chem. C} 2013 \textbf{117} 8941--8949
%
\bibitem{Teperik:2013a} Teperik T~V, Nordlander P, Aizpurua J
and Borisov A~G
\textit{Phys. Rev. Lett.} 2013 \textbf{110} 263901
%
\bibitem{Yan:2015a} Yan W, Wubs M and Mortensen N~A
\textit{Phys. Rev. Lett.} 2015 \textbf{115} 137403
%
\bibitem{Hohenester:2016} Hohenester U and Draxl C
\textit{Phys. Rev. B} 2016 \textbf{94} 165418
%
\bibitem{Toscano:2015} Toscano G, Straubel J, Kwiatkowski A,
Rockstuhl C, Evers F, Xu H, Mortensen N~A and Wubs M
\textit{Nature Commun.} 2015 \textbf{6} 7132
%
\bibitem{Yan:2015b} Yan W
\textit{Phys. Rev. B} 2015 \textbf{91} 115416
%
\bibitem{Ciraci:2016} Cirac\`{i} C and Della Sala F
\textit{Phys. Rev. B} 2016 \textbf{93} 205405
%
\bibitem{Raza:2013a} Raza S, Stenger N, Kadkhodazadeh S, Fischer S~V,
Kostesha N, Jauho A-P, Burrows A, Wubs M and Mortensen N~A
\textit{Nanophotonics} 2013 \textbf{2} 131--138
%
\bibitem{Kreibig:1970} Kreibig U and Zacharias P
\textit{Z. Physik} 1970 \textbf{231} 128--143
%
\bibitem{Kraus:1983}
Kraus W~A and Schatz G~C
\textit{J. Chem. Phys.} 1983 \textbf{79} 6130--6139
%
\bibitem{Kubo:1966} Kubo R \textit{Rep. Prog. Phys.} 1966 \textbf{29} 255
%
\bibitem{Mortensen:2014} Mortensen N~A, Raza S, Wubs M,
S{\o}ndergaard T and Bozhevolnyi S~I
\textit{Nature Commun.} 2014 \textbf{5} 3809
%
\bibitem{Apell:1984} Apell P, Monreal R and Flores F
\textit{Solid State Commun.} 1984 \textbf{52} 971--973
%
\bibitem{Ouyang:1992} Ouyang F, Batson P~E and Isaacson M
\textit{Phys. Rev. B} 1992 \textbf{46} 15421--15425
%
\bibitem{Baida:2009} Baida H, Billaud P, Marhaba S, Christofilos D,
Cottancin E, Crut A, Lerm\'{e} J, Maioli P, Pellarin M, Broyer M,
Del Fatti N, Vall\'{e}e F, S\'{a}nchez-Iglesias A, Pastoriza-Santos I
and Liz-Marz\'{a}n L~M
\textit{Nano Lett.} 2009 \textbf{9} 3463--3469
%
\bibitem{Lerme:2010} Lerm\'{e} J, Baida H, Bonnet C, Broyer M,
Cottancin E, Crut A, Maioli P, Del Fatti N, Vall\'{e}e F and Pellarin M
\textit{J. Phys. Chem. Lett.} 2010 \textbf{1}, 2922--2928
%
\bibitem{Monreal:2013} Monreal R~C, Antosiewicz T~J and Apell S~P
\textit{New J. Phys.} 2013 \textbf{15} 083044
%
\bibitem{Li:2013} Li X, Xiao D and Zhang Z
\textit{New J. Phys.} 2013 \textbf{15} 023011
%
\bibitem{Khurgin:2015a} Khurgin J~B
\textit{Faraday Discuss.} 2015 \textbf{178} 109--122
%
\bibitem{Uskov:2014} Uskov A~V, Protsenko I~E, Mortensen N~A
and O'Reilly E~P
\textit{Plasmonics} 2014 \textbf{9} 185--192
%
\bibitem{Kirakosyan:2016} Kirakosyan A~S, Stockman M~I and 
Shahbazyan T~V
\textit{Phys. Rev. B} 2016 \textbf{94} 155429
%
\bibitem{Shahbazyan:2016} Shahbazyan T~V
\textit{Phys. Rev. B} 2016 \textbf{94} 235431
%
\bibitem{Lerme:2017} Lerm\'{e} J, Bonnet C, Lebeault M-A,
Pellarin M and Cottancin E
\textit{J. Phys. Chem. C} 2017 \textbf{121} 5693--5708
%
\bibitem{Chandrasekhar:1943} Chandrasekhar S
\textit{Rev. Mod. Phys.} 1943 \textbf{15} 1--89
%
\bibitem{Landau-Lifshitz-Pitaevskii} Landau L~D, Lifshitz E~M
and Pitaevskii L~P 1984
\textit{Electrodynamics of Continuous Media} (Butterworth Heinemann)
%
\bibitem{Hanson:2010} Hanson G~W
\textit{IEEE Anten. Propag. Mag.} 2010 \textbf{52} 198--207
%
\bibitem{Iwamoto:1984} Iwamoto N, Krotscheck E and Pines D
\textit{Phys. Rev. B} 1984 \textbf{29} 3936--3951
%
\bibitem{Feibelman:1982} Feibelman P~J
\textit{Prog. Surf. Sci.} 1982 \textbf{12} 287--408
%
\bibitem{Kreibig:1985} Kreibig U and Genzel L
\textit{Surf. Sci.} 1985 \textbf{156} 678--700
%
\bibitem{Boardman:1982a} Boardman A~D 1982
\textit{Electromagnetic Surface Modes} (Wiley)
%
\bibitem{Tokatly:1999} Tokatly I and Pankratov O
\textit{Phys. Rev. B} 1999 \textbf{60} 15550--15553
%
\bibitem{Halevi:1995} Halevi P \textit{Phys. Rev. B} 1995 \textbf{51} 7497--7499
%
\bibitem{luo_prl111}
Luo Y, Fernandez-Dominguez A~I, Wiener A,
Maier S~A and Pendry J~B 2013
\textit{Phys. Rev. Lett.} \textbf{111} 093901
%
\bibitem{Ashcroft:1976} Ashcroft N~W and Mermin N~D
1976 \textit{Solid State Physics} (Harcourt)
%
\bibitem{Campos:2019} Campos A, Troc N, Cottancin E, Pellarin M,
Weissker H-C, Lerm\'{e} J, Kociak M and Hillenkamp M
\textit{Nature Phys} 2019 \textbf{15} 275--280
%
\bibitem{Rammer:1986} Rammer J and Smith H
\textit{Rev. Mod. Phys.} 1986 \textbf{58} 323--359
%
\bibitem{Wubs:2016} Wubs M and Mortensen N A Springer Series in Solid-State Sciences 2016 \textbf{185} 279--302
%
\bibitem{Tserkezis:2017b} Tserkezis C, Yan W, Hsieh W, Sun G,
Khurgin J~B, Wubs M and Mortensen N~A
\textit{Int. J. Mod. Phys. B} 2017 \textbf{31} 1740005
%
\bibitem{Raza:2015b} Raza S, Wubs M, Bozhevolnyi S~I
and Mortensen N~A
\textit{Opt. Lett.} 2015 \textbf{40} 839--842
%
\bibitem{Lindhard:1954} Lindhard J
\textit{Dan. Mat. Fys. Medd.} 1954 \textbf{28} 1--57
%
\bibitem{Yang:2019} Yang Y, Zhu D, Yan W, Agarwal A, Zheng M,
Joannopoulos J~D, Lalanne P, Christensen T, Berggren K~K and
Solja\v{c}i\'{c} M
\textit{Nature} 2019 \textbf{576} 248--252
%
\bibitem{Goncalves:2020} Gon\c{c}alves, P~A~D, Christensen, T, Rivera N,
Jauho A-P, Mortensen N~A and Solja\v{c}i\'{c} M
\textit{Nature Commun.} 2020 \textbf{11} 366
%
\bibitem{Tserkezis:2018} Tserkezis C, Ye\c{s}ilyurt A~T~M, Huang J-S and
Mortensen N~A
\textit{ACS Photonics} 2018 \textbf{5} 5017--5024
%
\bibitem{Christensen:2017} Christensen T, Yan W, Jauho A-P,
Solja\v{c}i\'{c} M and Mortensen N~A
\textit{Phys. Rev. Lett.} 2017 \textbf{118} 157402
%
\bibitem{David:2013} David C, Mortensen N~A and Christensen J
\textit{Sci. Rep.} 2013 \textbf{3} 2526
%
\bibitem{PrivateCommun} Gon\c{c}alves, P~A~D \textit{Private communication.} 2020.
%
\bibitem{wiener_nl12}
Wiener A, Fern\'{a}ndez-Dom\'{i}nguez A~I, Horsfield A~P,
Pendry J~B and Maier S~A 2012
\textit{Nano Lett.} \textbf{12} 3308--14
%
\bibitem{Bennett:1970} Bennett A~J
\textit{Phys. Rev. B}  1970 \textbf{1} 203--207
%
\bibitem{Tserkezis:2016a} Tserkezis C, Maack J~R, Liu Z,
Wubs M and Mortensen N~A
\textit{Sci. Rep.} 2016 \textbf{6} 28441
%
\bibitem{Tserkezis:2017} Tserkezis C, Wubs M and Mortensen N~A
\textit{Phys. Rev. B} 2017 \textbf{96} 085413
%
\bibitem{Grillet:2011}
Grillet N, Manchon D, Bertorelle F, Bonnet C, Broyer M,
Cottancin E, Lerm\'{e} J, Hillenkamp M and Pellarin M
\textit{ACS Nano} 2011 \textbf{5} 9450--9462
%
\bibitem{McMahon:2010a} McMahon J~M, Gray S~K and Schatz G~C
\textit{Phys. Rev. B} 2010 \textbf{82} 035423
%
\bibitem{Toscano:2012} Toscano G, Raza S, Jauho A-P,
Mortensen N~A and Wubs M
\textit{Opt. Express} 2012 \textbf{20} 4176--4188
%
\bibitem{Trugler:2017} Tr\"{u}gler A, Hohenester U and
Garc\'{i}a de Abajo F~J
\textit{Int. J. Mod. Phys. B} 2017 \textbf{31}, 1740007
%
\bibitem{Li:2017} Li L, Lanteri S, Mortensen N A, and Wubs M
\textit{Comp. Phys. Commun.} 2017 \textbf{219} 99
%
\bibitem{Doicu:2020} Doicu A, Eremin Y and Wriedt T
\textit{J. Quant. Spectr. Rad. Transf.} 2020 \textbf{242} 106756
%
\end{thebibliography}
\end{document}